\documentclass[a4paper,twocolumn,11pt]{quantumarticle} %
\pdfoutput=1
\usepackage[utf8]{inputenc}
\usepackage[english]{babel}
\usepackage[T1]{fontenc}
\usepackage{amsmath}
\usepackage{hyperref}

\usepackage{tikz}
\usepackage{lipsum}

\begin{document}

\title{Versatile quadrature antenna for precise control of large electron spin ensembles in diamond}

\author{Ruben Pellicer-Guridi}
\affiliation{Centro Física de Materiales, Donostia-San Sebastián, Spain.} 
\affiliation{Donostia International Physics Center, Donostia-San Sebastián, Spain.}
\orcid{0000-0001-9263-4921}
\author{Koen Custers}
\affiliation{Medical Image Analysis, Dept. Biomedical Engineering, Eindhoven University of Technology, Eindhoven, Netherlands}
\affiliation{Care\&Cure Lab of the Electromagnetics Group (EM4C\&C), Dept. Electrical Engineering, Eindhoven University of Technology, Eindhoven, The Netherlands.}
\orcid{0009-0002-9104-8714}
\author{Joseba Solozabal-Aldalur}
\affiliation{Centro Física de Materiales, Donostia-San Sebastián, Spain.} 
\author{Alexey Brodolin}
\affiliation{Donostia International Physics Center, Donostia-San Sebastián, Spain.}
\orcid{0000-0001-6398-7576}
\author{Jason T. Francis}
\affiliation{Centro Física de Materiales, Donostia-San Sebastián, Spain.} 
\orcid{0000-0003-1527-2407}
\author{Miguel Varga}
\affiliation{Centro Física de Materiales, Donostia-San Sebastián, Spain.} 
\orcid{0000-0002-8148-2990}
\author{Jorge Casanova}
\affiliation{EHU Quantum Center, Univeristy of the Basque Country UPV/EHU, Leioa, Spain.}
\affiliation{Department of Physical Chemistry, University of Basque Country UPV/EHU, Bilbao, Spain.}
\orcid{0000-0002-6110-671X}
\author{Margarethus M. Paulides}
\affiliation{Care\&Cure Lab of the Electromagnetics Group (EM4C\&C), Dept. Electrical Engineering, Eindhoven University of Technology, Eindhoven, The Netherlands.}
\affiliation{Dept. Radiotherapy, Cancer Institute, Erasmus University Medical Center, Rotterdam, The Netherlands.}
\orcid{0000-0002-5891-2139}
\author{Gabriel Molina-Terriza}
\affiliation{Centro Física de Materiales, Donostia-San Sebastián, Spain.}
\affiliation{IKERBASQUE, Basque Foundation for Science, Bilbao, Spain}
\orcid{0000-0002-7905-4938}

\maketitle

\begin{abstract}
We present an easily reproducible inexpensive microwave antenna that can generate a strong and homogeneous magnetic field of arbitrary polarization, which enables fast and coherent control of electron spins over a large volume. Unlike preceding works, we present a resonant antenna that maintains its resonant behaviour regardless of the proximity of other experimental hardware components. This robustness is crucial as it enables, amongst others, using microscope objectives with short working distances to perform wide field imaging/sensing with bulk diamonds. The antenna generates a magnetic field strength of 22.3 A/m for 1 W total driving power, which doubles the power efficiency compared with previously reported patch antenna designs. The magnetic field homogeneity in a volume of $0.3 \text{mm}^3$, $0.5 \text{mm}^3$ and $1 \text{mm}^3$ is within 6\%, 8\% and 13\%, respectively. The antenna has a full width at half maximum bandwidth of $\sim$160 MHz and its resonant frequency can be tuned over a 400 MHz range via four capacitors or varactors. The antenna has been tested and found to remain within safe handling temperatures during continuous-wave operation at 8 W. The files required to reproduce this antenna, which can be built on a standard and affordable double sided PCB, are provided open-source. This work facilitates a robust and versatile piece of instrumentation, being particularly appealing for applications such as high sensitivity magnetometry and wide field imaging/sensing with Nitrogen Vacancy centers.
\end{abstract}

\section{Introduction}
Nitrogen vacancy (NV) centers in diamond have emerged as a promising platform for quantum sensing due to their exceptional properties, such as long coherence times even at room temperature, high sensitivity, and nanoscale spatial resolution. They have been used to detect various physical quantities, including magnetic and electric fields, temperature, and strain \cite{Schirhagl2014Nitrogen-VacancyBiologyb}. Furthermore, their robustness, wide working temperature range and optical addressability make them attractive candidates for practical applications in diverse fields such as materials science, biomedicine, and fundamental physics \cite{Schirhagl2014Nitrogen-VacancyBiologyb, Budakian2023RoadmapImaging, Degen2016QuantumSensing}. 

NV centers are atomic-scale defects consisting of a nitrogen atom adjacent to a vacant lattice site, or vacancy, in the crystal structure of diamond. These defects can exist in three electronic states: neutral (NV$^{0}$), positively charged (NV$^{+}$), and negatively charged (NV$^{-}$). The latter is the most explored due to its remarkable optical and spin properties \cite{Doherty2013TheDiamond}. The NV$^{-}$ state is at the core of this work and will be referred to as NV center during the rest of this manuscript. In this configuration, a valence electron and an extra electron form a spin pair $S=1$. This free electron pair holds a spin triplet with state $m_{s} = 0, +1$ or $-1$. The energies of the spin states are sensitive to external influences. Additionally, contrary to most quantum systems, this spin has long coherence times at room temperature, making it one of the most prominent cryogenic free quantum platforms.

\begin{figure}[t]
    \centering
    \includegraphics[width=0.46\textwidth]{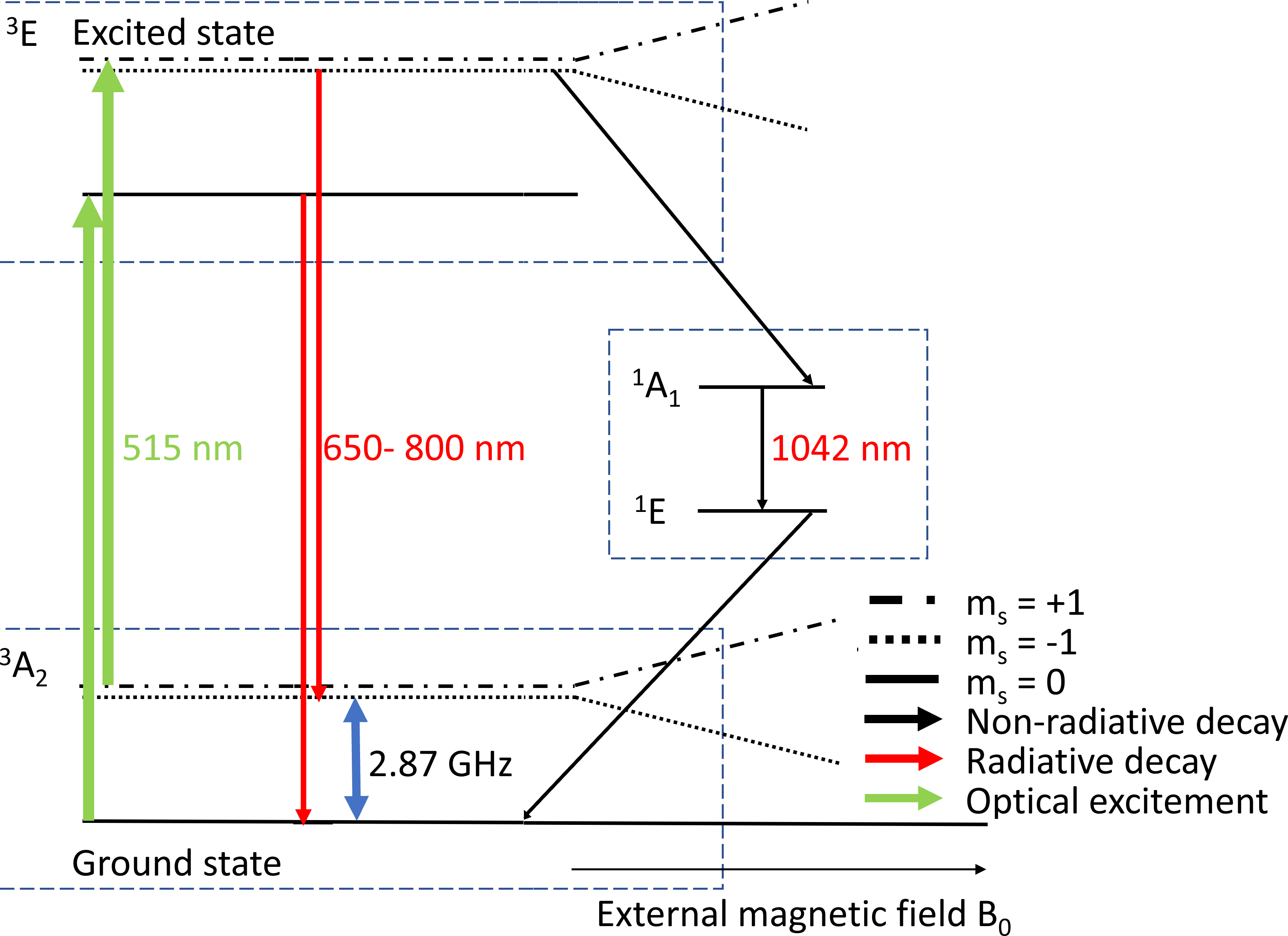}
    \caption{Simplified energy level structure and transitions of the NV$^{-}$ center. A green laser induces a spin conserving transition from the ground state to the excited state. $\sim$30 \% of the times the $m_{s} = \pm1$ spin states decay to the $m_{s} = 0$ state and do not emit red fluorescence. The $m_{s} = \pm1$ spins states are degenerate in the absence of an external magnetic field.}
    \label{fig:energy_levels}
\end{figure}

The initialization, control and readout of NV centers in diamond using laser and microwave (MW) techniques are fundamental processes in harnessing their quantum properties. A widespread NV center spin controlling approach is to use a green laser pulse to initialize the electron spin to its $m_{s} = 0$ state through a preferred decay pathway from the excited state, as visualized in \autoref{fig:energy_levels}. If this decay is spin conserving, which is the most likely case for the $m_{s} = 0 $ state, a photon is emitted predominantly in the 600-800 nm spectral range. On the other hand, the excited $m_{s} = \pm1$ states may not decay back to the same spin state, but they can decay to the ground $m_{s} = 0 $ state through another path with a probability of around $30\%$, emitting photons with 1042 nm wavelength \cite{Acosta2010OpticalDiamond}. Thereby, this pathway allows for both initialising and inferring the state of NV centers. Thanks to the Zeeman effect, in the presence of an external magnetic field the $m_{s} = \pm1$ spin states are energetically separated. This energy gap between the electron spin states enables the manipulation of spin states into an arbitrary superposition with $m_{s} = 0, +1$ and $-1$ by means of microwave (MW) fields. The energy gap between the three electron spin states depends on the external magnetic field. As a consequence, the necessary microwave frequency to excite the $m_{s} = 0 \Longleftrightarrow \pm1$ transitions also changes, as shown in \autoref{fig:energy_levels}. 

By considering only the effects of the external magnetic field, the electron spin Hamiltonian can be simplified to

\begin{equation}
    \label{eq:hamiltonian}
    H = DS_{z}^2-\gamma_{e}\vec{B} \cdot \hat{S},
\end{equation}

where $D = 2.87$ GHz is the zero-field splitting parameter, $\hat{S}$ is the vector of spin operators and $\gamma_{e} = 2.8$ MHz/G is the gyro-magnetic ratio of the electron \cite{Gali2019AbDiamond}. Another effect that influences the NV energy gap is the Stark effect under the presence of an electric field \cite{Gali2019AbDiamond}. Strain modifies the crystal lattice structure of the diamond, which in turn also alters the energy levels and transition frequencies of the NV center \cite{Gali2019AbDiamond}. Fluctuations in temperature variations can also cause shifts in the energy levels and transition frequencies of NV centers over time \cite{Gali2019AbDiamond}.

Working near the degeneracy conditions of the $m_{s} = \pm1$ states, e.g., without an externally provided bias magnetic field \cite{Lenz2021MagneticCenter}, hinders spin control as linearly polarized MW fields cannot address the $m_{s} = \pm1$ states individually without introducing strong constraints in pulse sequences \cite{Vetter2022Zero-Centers}. Conversely, circularly polarized MW fields can act selectively on $m_{s} = +1$ or $m_{s} = -1$ states depending on the direction of the polarization \cite{Alegre2007polarization-selectiveDiamond}. 

In this context, the design and the characteristics of the microwave antenna used for NV center experiments play a crucial role in achieving efficient and precise manipulation of the NV center's spin states. Factors such as the antenna geometry, polarization, and power distribution should be optimized to maximize the coupling efficiency between the antenna and the NV center, ensuring effective spin manipulation. Large ensembles of NV centers can be advantageous for several applications, i.e. vector magnetometry determining both direction and magnitude of magnetic ﬁelds simultaneously \cite{Schloss2018SimultaneousSpins}, parallel sensing in different spatial locations for imaging applications \cite{LeSage2013OpticalCells}, improving measurement sensitivity by averaging the responses of multiple NV centers \cite{Taylor2008High-sensitivityResolution}, NMR spectroscopy with NV ensembles \cite{Glenn2018High-resolutionSensor}, etc. However, ensemble sensing requires addressing multiple NV centers simultaneously with the same microwave field. This is a technical challenge requiring microwave antenna designs with reduced inhomogeneity over a large volume in the diamond. 

When employing single NV centers it is a common practice to use a simple straight wire or a loop as source for the MW fields, which can enable very fast spin control due to the high magnetic fields they generate when placed in very close proximity to the NV centers. However, these antennas are not suitable to control NV ensembles as they generate very different MW conditions for distinct NVs because the MW fields they generate possess large spatial variations. Furthermore, it is more difficult to achieve orthogonality between the MW magnetic field and that of the external static one at the location of the NV center \cite{Abe2018Tutorial:IN}, \cite{Bucher2019QuantumSpectroscopy}. 

A diversity of microwave antennas have been proposed to address these homogeneity limitations. Many of the existing works limit to generating linearly polarized fields \cite{Wang2020IntegratedSensorsb,Opaluch2021OptimizedApplications, Sasaki2016BroadbandDiamond, Bayat2014EfficientResonators, Chen2018Large-areaDiamond}. 

Several approaches have been proposed capable of generating circularly polarized MW magnetic fields but with reduced field homogeneity. A pair of wires aligned either orthogonally or perpendicularly has been used to generate circularly polarized magnetic fields. Such designs based on thin straight wires are placed few a $\mu$m away from the NV centers to the achieve high magnetic field ratios, and suffer very high spatial inhomogeneity \cite{London2014StrongFields}, \cite{Mrozek2015CircularlyDiamonds}. Wide bandwidth microstrip resonators have been proposed on standard PCBs \cite{MayerAlegre2007Microstrippolarization} and on optically transparent conductive materials \cite{Staacke2020MethodManipulation}, with the former blocking the optical access from one face of the diamond and the latter enhancing this access at the expense of requiring non-standard manufacturing processes and materials. These microstrip designs improve the homogeneity in the plane parallel to the surface. However, they still present large field variations along the direction orthogonal to the microstrip, constraining the effective volume homogeneity \cite{Staacke2020MethodManipulation}. 

Resonant antennas can offer a solution to covering a large volume with a strong and homogeneous field \cite{Sasaki2016BroadbandDiamond, Herrmann2016polarization-Diamond, Yaroshenko2020CircularlyDiamond}. For example, a dielectric ceramic resonator with tailored high dielectric constant has been proposed which achieves a Rabi frequency of 3.9 MHz/$\sqrt{\text{W}}$, a full width at half maximum bandwidth (FWHM) of 140 MHz at 2.67 GHz, and an excellent theoretical inhomogeneity of 4\% (reported as standard deviation) over a cylindrical volume of $5.9 \text{mm}^3$ (measured $\sim$ 7.3\% over $0.45 \text{mm}^3$)\cite{Yaroshenko2020CircularlyDiamond}. This design allows ample optical access from both axial sides, but is limited by design to a fixed resonant frequency and a fixed polarity of the circular polarization. Moreover, its robustness against detuning due to nearby elements has not been reported. This last caveat is common in resonant antennas \cite{Herrmann2016polarization-Diamond} \cite{Bayat2014EfficientResonators}. A typical example is given in \cite{Herrmann2016polarization-Diamond}, showing a strong shift in the resonant frequency of $\sim$70 MHz due to the position of an objective lens. Yet, this planar antenna concept has very appealing properties such as an actively re-tunable resonant frequency and a seamless reproducibility with standard PCB materials and manufacturing processes \cite{Herrmann2016polarization-Diamond}. This previous reference work featured 165 MHz FWHM, and enabled a Rabi frequency of 0.8 MHz/$\sqrt{\text{W}}$ \cite{Herrmann2016polarization-Diamond}. To the best of our knowledge, no resonant microwave antenna capable of generating circular polarization has been reported that is robust against external hardware components commonly employed with NV center setups.

Here we present an efficient quadrature microwave antenna that maintains its calibrated resonant behaviour regardless the configuration of its neighbouring setup components. This largely reduces setup constraints otherwise introduced by antennas whose behaviour is dependent on nearby elements \cite{Herrmann2016polarization-Diamond}, and enables dynamic configurations such as scanning movements of microscope objectives. Furthermore, the design proposed here complements the benefits of the design of Herrmann et al. \cite{Herrmann2016polarization-Diamond} with enhanced optical access, power efficiency, and reproducibility. The antenna has been tested numerically and experimentally to validate its reflection and transmission coefficients, magnetic field amplitude and homogeneity, and power dissipation. Being aware that MW antennas are one of the few elements not available off-the-shelf, we offer open-source the files required to reproduce the antenna and accelerate scientific progress and reproducibility  \cite{repoQuadMWPatchAntenna2024}. We also show that further performance enhancement could be possible using a PCB substrate with higher dielectric constant at the expense of reducing the bandwidth.

\section{Design materials and methods}
\subsection{Antenna Design}
The antenna was designed to generate strong and homogeneous circularly polarized fields while providing ample optical and mechanical access and fabrication simplicity. The planar geometry of the circular patch antenna together with the through hole to host the diamond facilitates easy access space on both sides of the PCB so other setup components, like microscope objectives with short working distances or samples to be analyzed, can be brought in close proximity to the diamond. This design also offers more flexibility when designing optical path configurations because optical pumping and fluorescence collection can be performed from both faces of the diamond. The resonant frequency of the prototype was designed to be 2.86 GHz. However, to enable efficient control of the $m_{s} = 0 \Longleftrightarrow -1$ transition, the resonant frequency can be tuned to lower frequencies by adding capacitors to the antenna. As mentioned earlier, two important design constraints are achieving a high field strength and high homogeneity over a large volume in the diamond. Importantly, to harvest the highest signal to noise ratio, the location and shape of this volume should be such that it can be optically addressed in a manner that the NVs contained in the optimally controllable volume are all interrogate at once, while minimizing the fluorescence from NVs not contained in this volume.

An electromagnetic field simulation software (CST Studio Suite 2023, Dassault Systèmes, Vélizy-Villacoublay, France) is used to optimize antenna design parameters. An initial geometry is estimated from analytical expressions for circular patch antennas derived in \cite{Balanis2016AntennaDesign}. The antenna is sized to match its first resonant mode $\text{TM}_{110}$ to the desired MW frequency. This mode induces the most homogeneous and the largest magnetic field that is achievable with this type of circular antenna \cite{Balanis2016AntennaDesign}. Its corresponding resonant frequency is estimated by: 
\begin{equation}
    \label{eq:circular_antenna}
   (f_{r})_{110} = \frac{1.8412c}{2 \pi a \sqrt{ \epsilon_{r}}},
\end{equation}
where $c$ is the speed of light, $a$ is the radius of the antenna and $\epsilon_{r}$ is the permittivity in the dielectric material, also called the dielectric constant. The $\text{TM}_{110}$ mode presents the highest magnetic field at the centre of the antenna. Therefore, the positioning for the diamond has to be chosen accordingly. Other higher frequency modes could be excited but they will result in a lower and less homogeneous magnetic field. The antenna is designed with a square hole, so that it can host a diamond as large as 3 x 3 x 0.5 mm$^3$, which covers most commercially available single crystal samples. The simulations also include the diamond material with the aforementioned size, a dielectric constant $\epsilon_{r}$  of 5.7, and placed in the center of the hole.

\begin{figure}
     \centering
     \includegraphics[width=0.48\textwidth]{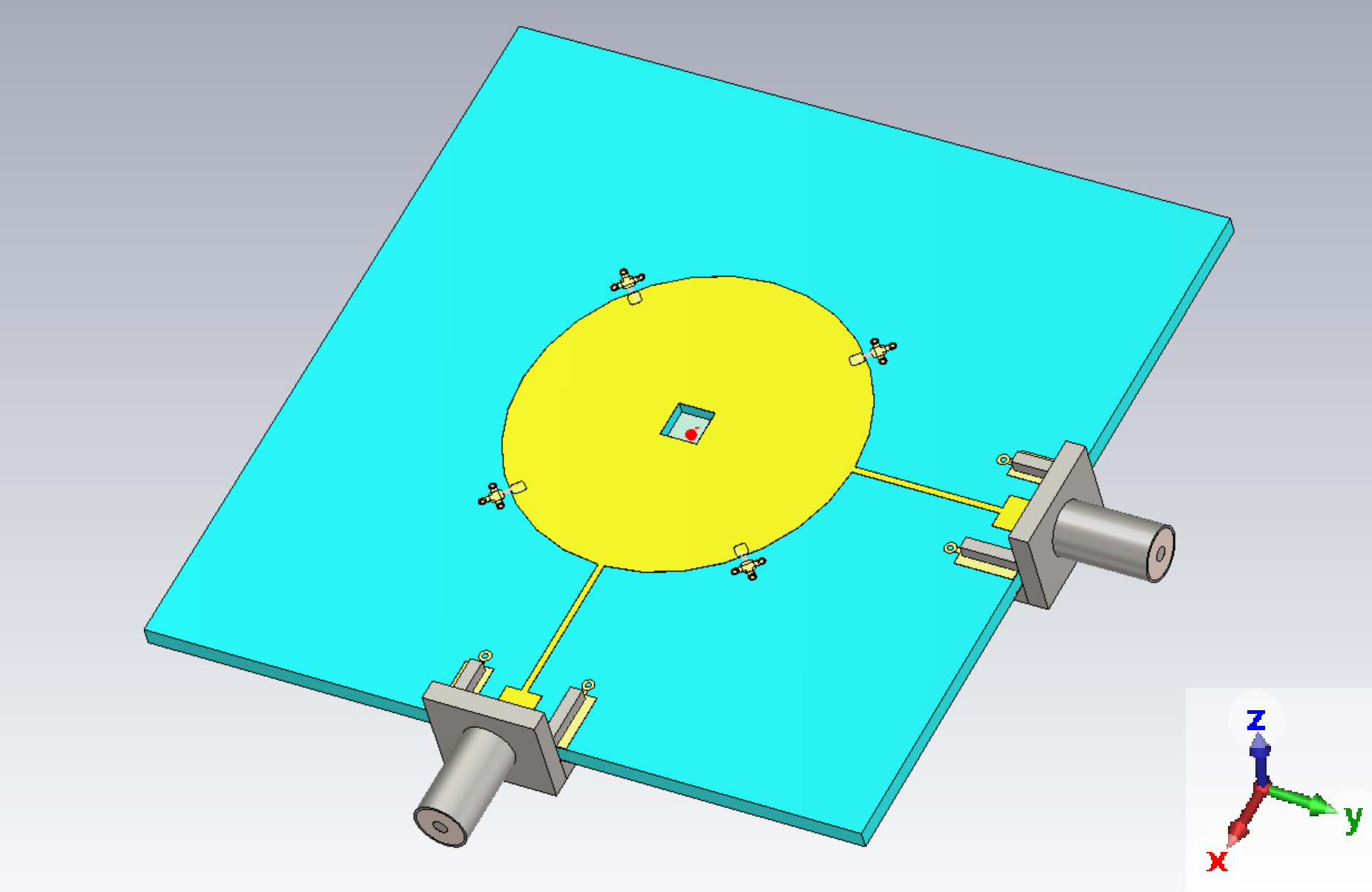}
     \label{fig:antenna_sideview}
     \caption{Simulated antenna for MW excitation of the NV center. It consists of a circular patch with the two feeding ports. The PCB is double-sided with a bottom copper plane covering the whole PCB area, excluding a centered square through hole that contains an embedded bulk diamond.}
     \label{fig:antenna_simulated}
\end{figure}

Due to its common availability, we have chosen FR-4 as PCB substrate, which has an $\epsilon_{r}$ of 4.4. According to \autoref{eq:circular_antenna}, by changing the dielectric constant, the size of the patch of the antenna will change proportionally with the factor of $\sqrt{\epsilon_{r}}$. Reducing substrate thickness also increases the magnetic field, but considerably penalizes the homogeneity along the Z direction. Coaxial SMA connectors are used for signal launch to the PCB, providing a reproducible performance and a robust connection. These connectors also facilitate maintaining a 50 $\Omega$ impedance of the transmission chain, thereby minimizing power reflections. A quarter wavelength transmission line is used to match the impedance of the circular patch antenna to the 50 $\Omega$ of the transmission chain. This enables efficient power transfer and high signal integrity. The feeding ports are connected to the outer edges of the circular patch, allowing surface currents to run through the patch without disturbance. This configuration ensures that the strongest magnetic field would be created in the middle of the patch. The antenna is visualized in \autoref{fig:antenna_simulated}.

\subsection{Simulation results}

\begin{figure}[h]
    \centering
    \includegraphics[width=0.4\textwidth]{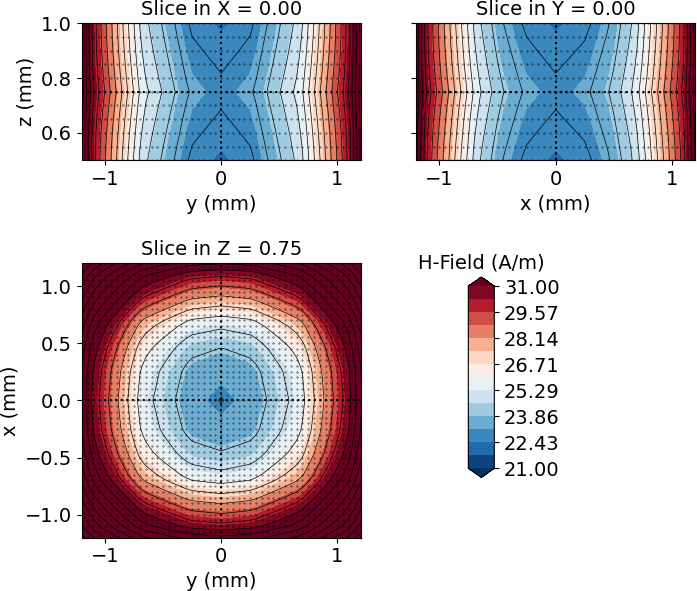}
    \caption{Simulated H-field for circular polarization mode. Cross-sections of the H-field on x, y, and z planes in the middle of the diamond. The values correspond to RMS values for 0.5 W per port fed with a $90^\circ$ dephase between them.}
    \label{fig:H_map_simul-CircPol}
\end{figure}

\begin{figure}[h]
     \centering
     \includegraphics[width=0.4\textwidth]{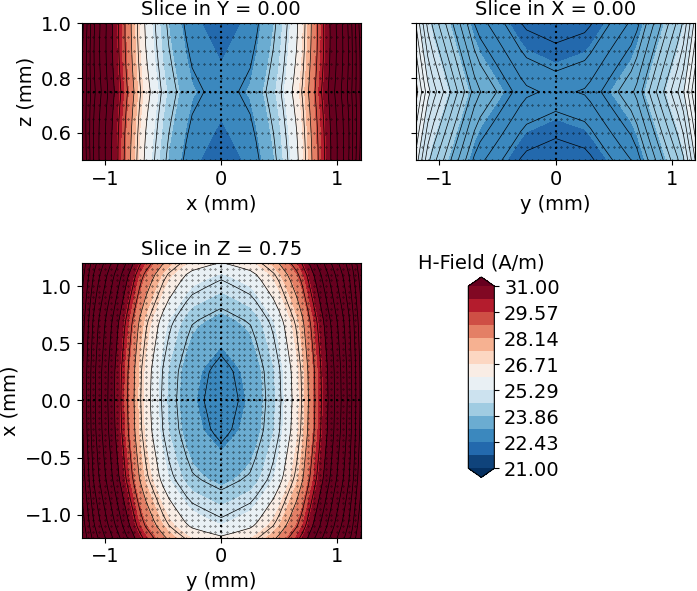}
     \caption{Simulated H-field for linear polarization mode. Cross-sections of the H-field on x, y, and z planes in the middle of the PCB and diamond. The values correspond to RMS values for 1 W power on a single port fed along the x=0 axis.}
     \label{fig:H_map_simul-LinearPol}
     \hfill
\end{figure}

In \autoref{fig:H_map_simul-CircPol} and \autoref{fig:H_map_simul-LinearPol} we present the simulations performed under different input polarization conditions. According to the simulations shown in \autoref{fig:H_map_simul-CircPol}, for the circular polarization mode, the inhomogeneity is within 13\% in a 1 mm$^3$ cylinder (500 $\mu$m height  and 800  $\mu$m  radius). For smaller volumes of 0.5 mm$^3$ and 0.3 mm$^3$, the inhomogeneity is within 8\% and 6\% respectively. In linear polarization modes ( \autoref{fig:H_map_simul-LinearPol}) the inhomogeneity is lower in the xy plane direction orthogonal to the axis of the feeding port, e.g., in the figure the feeding port is along the x = 0 axis, leading the y = 0 axis to have lower inhomogeneity. The field strength is most homogeneous near the center of the diamond in the xy plane, which also holds a local minimum for this plane. In the z axis, the field is the strongest and most homogeneous around the center of the PCB depth.

The spatial distribution of the magnetic field strength along the X and Y dimensions is elliptical, as can be seen in \autoref{fig:H_map_simul-LinearPol}. This ellipticity is similar on both ports but rotated by $90^\circ$ according to the spatial location of each port. The combination of the contribution of both ports generates a more circular RMS magnetic field as shown in \autoref{fig:H_map_simul-CircPol}. 

The RMS magnetic field values in the center of the region of interest is $22.8$ A/m RMS for 1 W total driving power in both driving modes, as visualized in \autoref{fig:H_map_simul-LinearPol} and \autoref{fig:H_map_simul-CircPol}.

\section{Experimental validation}

\subsection{Antenna fabrication and electrical characterization}
\begin{figure}[h]
    \centering
    \includegraphics[width=0.4\textwidth]{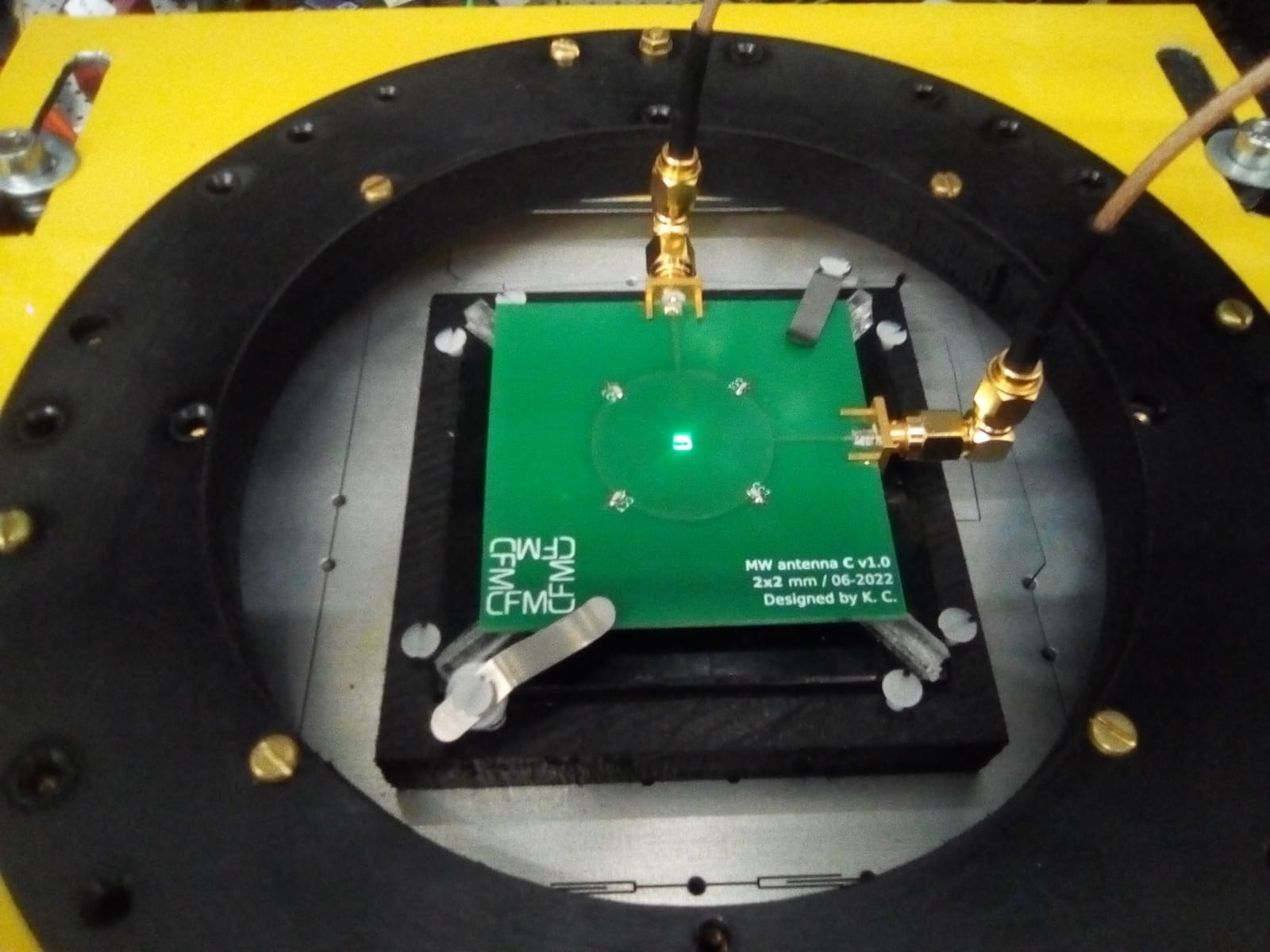}
    \caption{Photograph of the MW antenna during a measurement.}
    \label{fig:MW_antenna_during_measurement}
\end{figure}

An antenna designed to host a 3 x 3 x 0.5 mm$^3$  diamond size is built on a double sided standard FR-4 PCB (\autoref{fig:MW_antenna_during_measurement}). A commercially available diamond sample containing a high density ($\sim$4.5 ppm) of NV centers (DNV™ B14, Element Six Ltd, UK) is embedded in the hole of the MW antenna. Two high frequency SMA PCB edge launch connectors (ref. 142-9701-811, Cinch Connectivity Solutions, USA) are soldered onto the PCB, which are non-magnetic to minimize any distortion of DC fields when performing magnetic field measurements.

\begin{figure}[h]
     \centering
         \centering
         \includegraphics[width=0.46\textwidth]{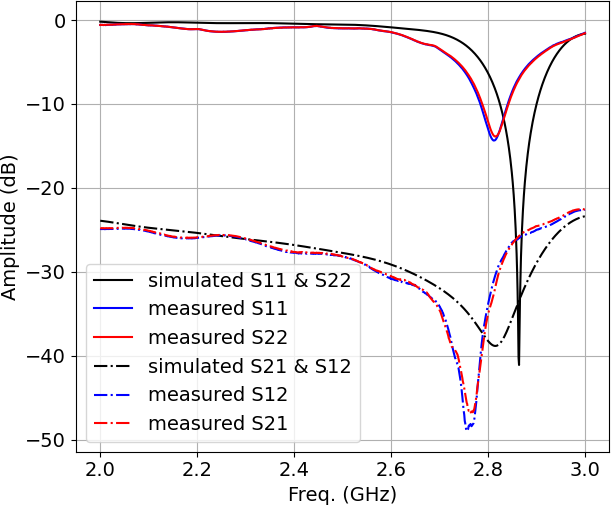}
        \caption{Plots of the measured and simulated reflection and transmission parameters of the antenna. The resonant frequency is 55 MHz lower in the manufactured antenna (2.809 GHz) as compared with the simulations (2.864 GHz). The reflection parameters of the manufactured antenna are -14 dB as compared with the theoretical -41 dB of the simulations. The transmission parameters are below -22 dB in all cases. Both ports of the manufactured antenna show a very similar response (blue and red lines).}
        \label{fig:S_parameters_sim_meas}
\end{figure}

To drive the antenna in a linearly polarized mode, a MW drive signal is fed through one of the ports. To generate the circularly polarized modes, the MW drive power signal is split into two $90^\circ$ phase shifted signals with a $90^\circ$ phase shifting power splitter (ZX10Q-2-34-S+, Mini-circuits, NY, USA), outputs of which are connected to the two ports of the antenna. Right or left circularly polarized fields are generated by swapping the order of the cables at the power splitter. A vector network analyser (Fieldfox N9912A, Keysight, USA) is used to measure the reflection (S11 \& S22)  and transmission (S21 \& S12) coefficients of the antenna, and to calibrate the power delivered to the antenna. The results of these measurements can be observed in \autoref{fig:S_parameters_sim_meas}. Overall, the experimental and simulated S parameters agree qualitatively. However, a small shift in the expected resonant frequency of the antenna and a worsening of the reflection coefficient can be observed. The simulated and measured response time of the antenna is 2 ns, which corresponds to a quality factor of 17.6 and a Full width at half maximum (FWHM) of 160 MHz\cite{QfactorMeasurement2020} and is in good agreement with the frequency response reflected in  \autoref{fig:S_parameters_sim_meas}.

\begin{figure}[h]
    \centering
    \includegraphics[width=0.46\textwidth]{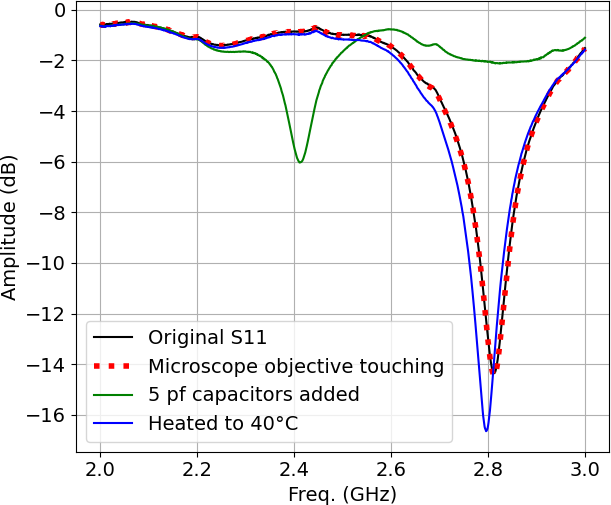}
    \caption{Measured reflection coefficients of the antenna under different conditions. The black dotted curve shows the behaviour of the antenna under standard operation conditions. The red line shows the antenna when the microscope objective is placed against the ground plane. The overlap between the dotted black line and the red line shows no change in the resonant frequency, confirming the robustness of the antenna to external hardware components approached from the ground plane. The green line visualizes the tunability of the antenna via lumped capacitors, in this case of 5 $pF$. The blue line shows the effect of temperature on the behaviour of the antenna, where a change from 22$^{\circ}C$ to 40$^{\circ}C$ decreases the resonant frequency by 16 MHz.}
    \label{fig:s_parameters_ext_conditions}
\end{figure}

 We have also tested the antenna under different conditions, to observe the resonant behaviour of the antenna (\autoref{fig:s_parameters_ext_conditions}). For example, we have added SMD capacitors of 5 pF with case code 0603 (1608 metric) to the antenna for lowering down the resonance frequency of the MW antenna from 2.81 to 2.41 GHz. This configuration reduces the resonant frequency by 400 MHz, which is equivalent to a dynamic range of the DC external magnetic field of almost 15 mT. It should be noted that the efficiency of the antenna is reduded by half when capacitively tuned to 2.41 GHz. 
 
 One important feature of this antenna is that it is very stable under changes of external conditions. For example, in \autoref{fig:s_parameters_ext_conditions} it can be observed that the features of the antenna do not change at all even if the optical microscope is in contact with the antenna. Also, changes in the temperature have a very small  effect on the resonant behaviour of the antenna.

\subsection{Setup} 
\begin{figure}[h]
    \centering
    \includegraphics[width=0.46\textwidth]{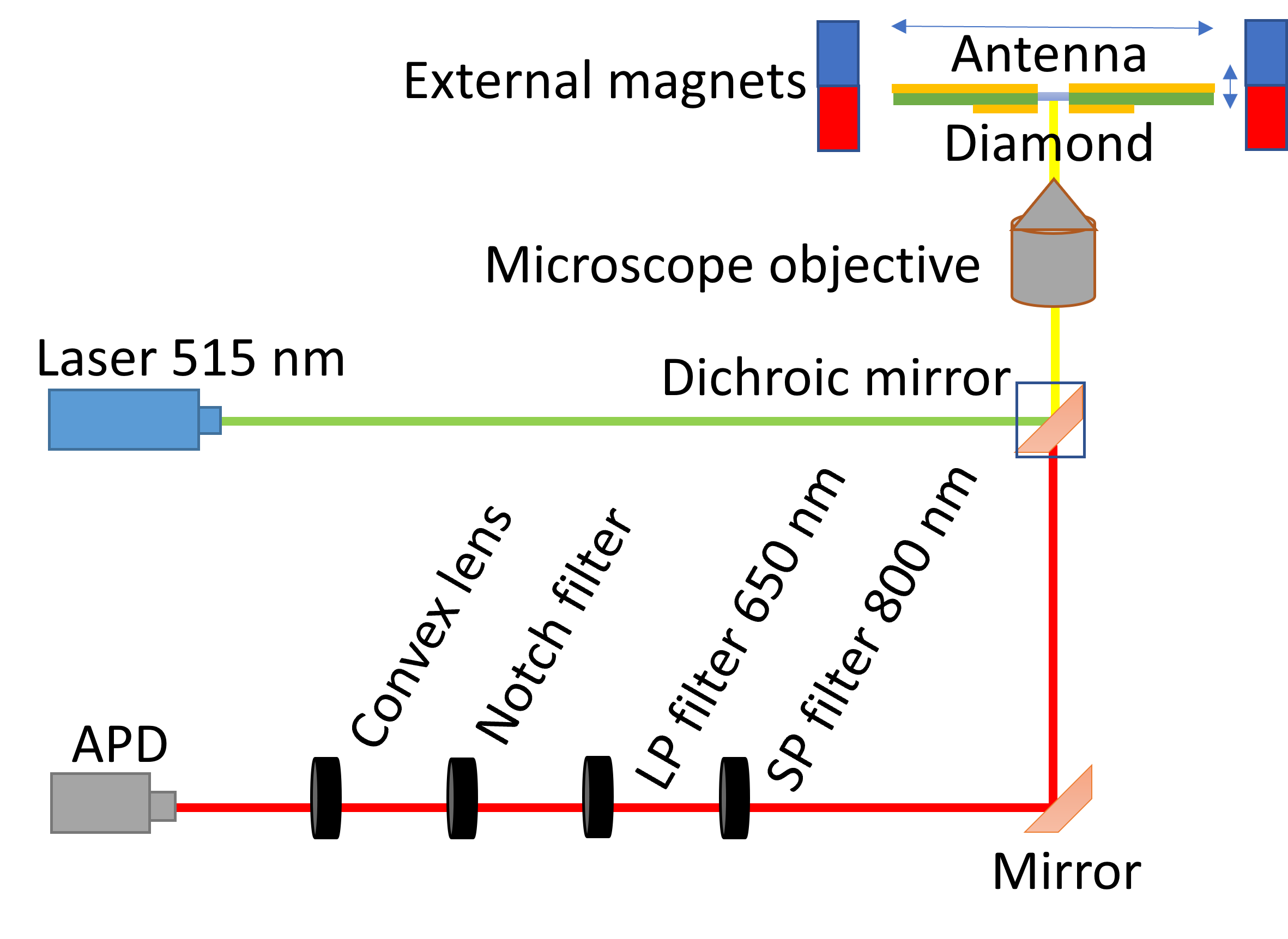}
    \caption{Overview of the confocal microscope employed to evaluate the MW antenna, with emphasis on the optical components. Shown are the pumping path (green laser), the confocal reading path (red and yellow), the permanent magnets array and the MW antenna.}
    \label{fig:setup}
\end{figure}

An optical setup is built to evaluate the manufactured MW antenna prototype. The setup allows to sweep through the volume of the diamond initializing and reading out the electron spin of the NV centers. The simplified schematic of the setup can be found in \autoref{fig:setup}. A green laser (iBeam-smart, 515 nm, Toptica, Germany) is used to excite the NV centers. The position of the antenna is determined by a micro-positioner that is programmatically controlled in the $x$,$y$ and $z$-axes (X-NA08A25-E09 Zaber, Canada  and SPM-MZ, Mad Citylab, Switzerland ). The excitation green laser beam is focused onto the sample with a microscope objective (LU Plan ELWD 50x, NA 0.55, WD 10.1, Nikon, Japan), which also collects the fluorescence from the diamond. This fluorescence is separated from the green excitation light by a dichroic mirror, a 515 nm notch filter, a 650 nm long pass (LP) filter and a 800 nm short pass (SP) filter. The intensity of this red fluorescence is measured by an avalanche photo-diode (A-CUBE-S3000-03, Laser Components, Germany).

The microwave signal is generated via a vector signal generator (SMJ100A, Rohde \& Schwarz, Germany) and a power amplifier $\sim$43 dB (ZHL-16W-43-S+, Mini-circuits, USA). A fast solid state switch (ZASWA-2-50DRA+, Mini-circuits, USA) is placed between the signal generator and the power amplifier when short pulses are needed.

\subsection{External magnetic field}
An external static magnetic field highly stable and of low spatial inhomogeneity is generated via a ring with permanent magnets placed at a radial distance of 80 mm. This ring can host up to 32 square magnets of 12 mm length. NdFeB magnets of grade N48 are employed. This external magnetic field points along the $z$ direction, i.e., orthogonal to the PCB of the MW antenna. The applied external magnetic field is fine-tuned for the Larmor frequency of the $m_{s} = 0 \Longleftrightarrow -1$ transition corresponding to $f \cong 2.81$ GHz following \autoref{eq:hamiltonian}.

\subsection{Acquisition sequences}
\begin{figure}
    \centering
    \includegraphics[width=0.46\textwidth]{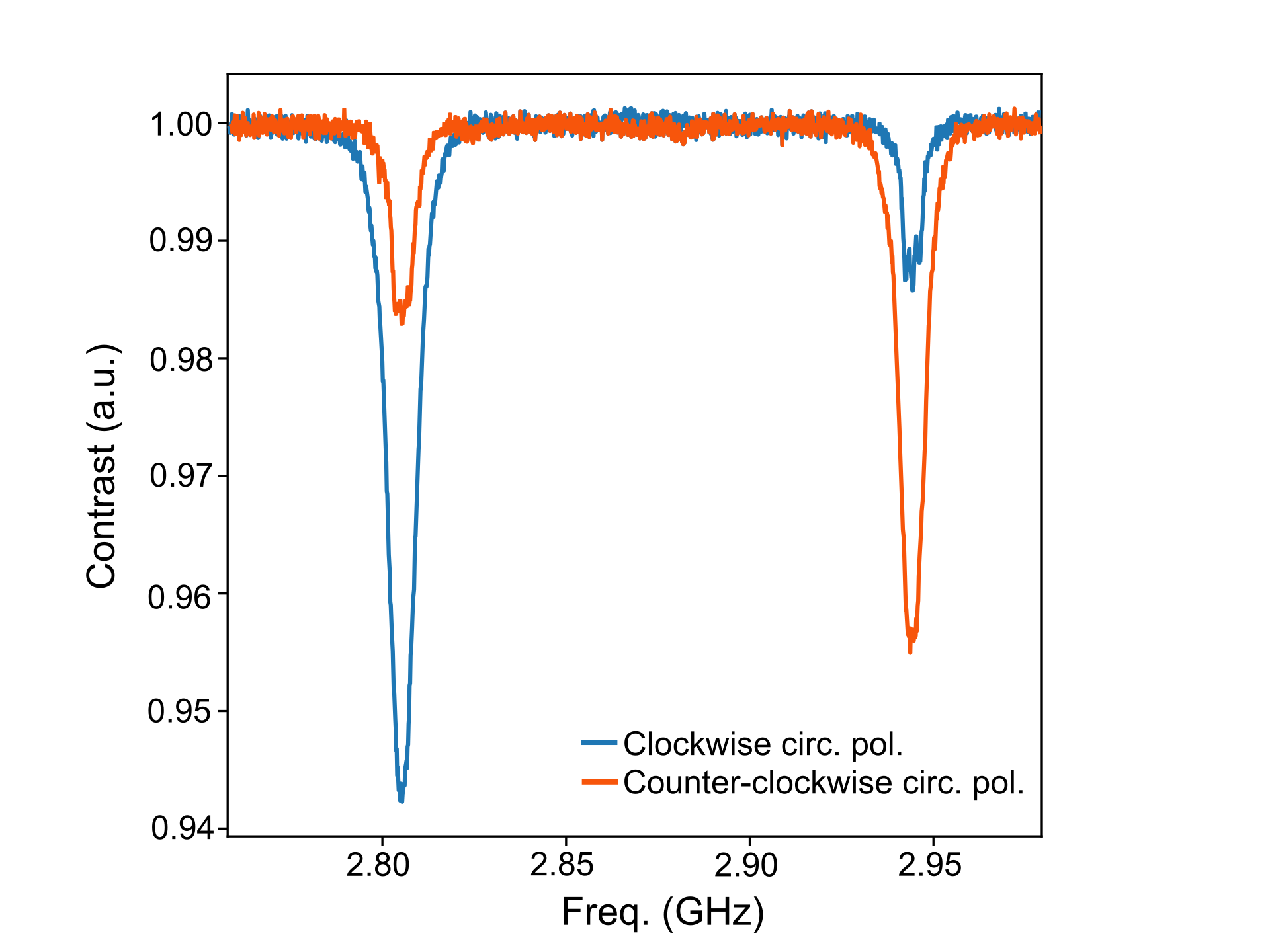}
    \caption{Plot of the ESR-experiment with both circular polarization directions. The blue (orange) line corresponds to the counter-clockwise (clockwise) polarization matching the $m_{s} = 0 \Longleftrightarrow -1 $ ($m_{s} = 0 \Longleftrightarrow +1 $) electron spin transition. A simple power splitter was used to feed both ports from a single power amplifier and no further efforts were made to tune the feeding powers and dephase. This far from ideal feed introduced a considerable ellipticity, which can be inferred from the presence of the second smaller peaks. The contrast of the $m_{s} = 0 \Longleftrightarrow +1 $ transition at 2.935 GHz is lower than the $m_{s} = 0 \Longleftrightarrow -1 $ counterpart at 2.805 GHz. This difference is due to the MW antenna being tuned to be resonant at 2.81 GHz and, as consequence, it is more efficient driving the $m_{s} = 0 \Longleftrightarrow +1 $ transition.}
    \label{fig:esr}
\end{figure}

An Electron Spin Resonance (ESR) experiment is performed where the change in the fluorescence due to microwave coupling to the electron spin states is measured. A typical example can be seen in \autoref{fig:esr}. It shows the fluorescence dip at $B_{z} \cong 2$ mT external magnetic field normal to the PCB and, since the crystallographic orientation of the used diamond is {100}, this field is aligned with the same angle to all NV axis (54.7$^{\circ}$).  Thereby, all the NVs will respond to the same MW frequency. The dips on the left correspond to the $m_{s} = 0 \Longleftrightarrow -1 $ electron spin transition and the dips on the right corresponds to the $m_{s} = 0 \Longleftrightarrow 1 $ transition. This experiment is used to confirm the generation of circular polarization and to adjust the MW frequency to be employed in following Rabi experiments. In \autoref{fig:esr} the hyperfine coupling between electron and nuclear (N$^{14}$) spins can be inferred from the three dents within the smaller dips. These dents are not visible on the longer peaks because the employed ESR sequence uses continuous-wave MW fields, which is known to produce power broadening if the transitions are strongly driven, i.e., the corresponding circularly polarized MW fields generated by the antenna are higher at those frequencies.

The strength and polarity of the MW field generated by the antenna are measured experimentally with a Rabi experiment. The sequence consists of MW pulses with different time durations. Subsequently, the oscillation frequency of the obtained photoluminescence is analyzed. The Rabi frequency is estimated by fitting the signal to an exponentially decaying oscillation according to
\begin{equation}
    \label{eq:rabi}
   S = A \sin (2 \pi\, \Omega\, \tau + \theta)\,    \exp\left(-\frac{\tau}{T_2}\right) + B\tau + C,
\end{equation}
where $\Omega$ is the Rabi frequency, $A$ is the amplitude of the signal, $\tau$ is the MW pulse duration, $\theta$ is a phase offset factor, $T_2$ is the decay ratio, $B$ is a linear drift and $C$ a DC offset.

The Rabi frequency is proportional to the MW magnetic field generated by the antenna \cite{Yaroshenko2020CircularlyDiamond},
\begin{equation}
    \label{eq:rabi2Hfield}
   \Omega = 2 \gamma_{e} | \langle 0|^{S}_{x}|1 \rangle B| = \gamma_{e} \mu_{0} \mu_{r} H_{rms},
\end{equation} where an additional $\sqrt{2}$ is multiplied in the case of circular polarization. Also, $H_{rms} \propto \sqrt{P_{ant}}$, the square root of the input power to the antenna.

\begin{figure}[h]
    \centering
    \includegraphics[width=0.46\textwidth]{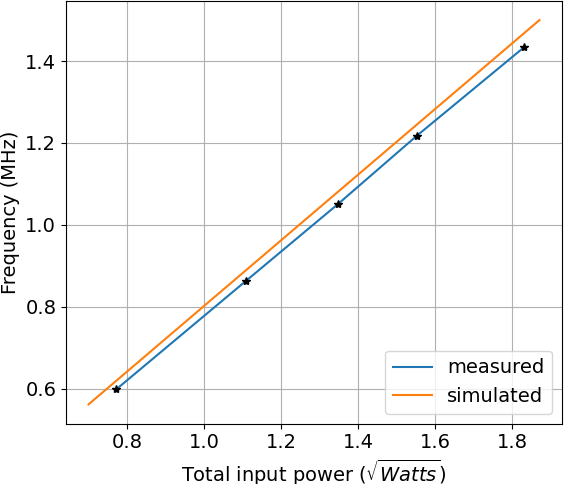}
    \caption{Measured dependence of the Rabi frequency to the total input MW power for linearly polarized configuration. The Rabi frequency increases proportionally to the square root of the power, validating the linear behaviour of the MW antenna.}
    \label{fig:rabi_freq_power}
\end{figure}

\autoref{fig:rabi_freq_power} shows that the Rabi oscillations mediated by the antenna follow this theoretical linear dependency between the square root of the applied power and the Rabi frequency. This verifies that the antenna works in a linear regime, which is important to calibrate the MW power being delivered to the diamond sample.

\subsection{Experimental characterization of the inhomogeneities of the antenna}
The distribution of Rabi frequencies across different planes (xy, xz, and yz) is analyzed using both circular and linear polarization. This spatial distribution is characterized stepping the position of the antenna and the diamond with the micropositioner, sweeping the spot being interrogated in the diamond. For circular polarization, each port of the antenna was fed with a power of 2.8 W, while for linear polarization, 3.8 W was applied to a single port. In the former, the power splitter fed to each port 1.3 W and 1.5 W.

\begin{figure}[h]
    \centering
    \includegraphics[width=0.46\textwidth]{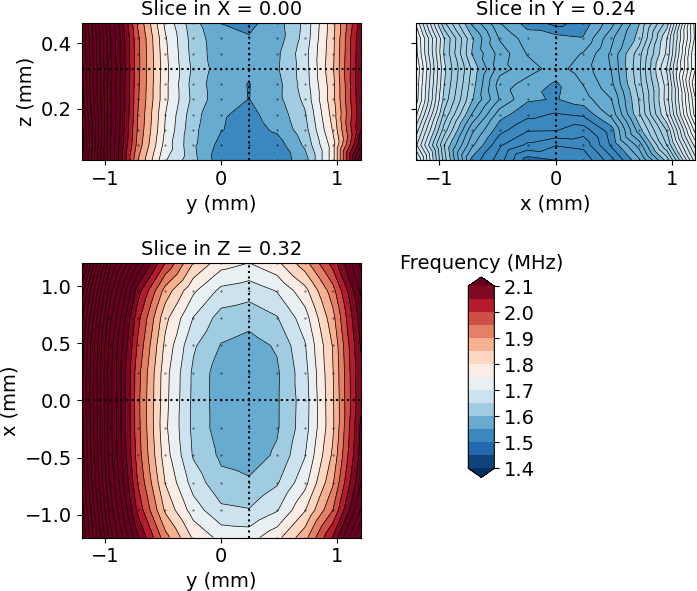}
    \caption{Measured Rabi frequency distribution maps for a linear polarized excited antenna at 3.8 W input power on the port feeding along the x=0 axis. The excitation MW frequency was 2.81 GHz, corresponding to the $m_{s} = 0 \Longleftrightarrow -1$ transition.}
    \label{fig:Rabi_map-LinearPol}
\end{figure}

\begin{figure}
    \centering
    \includegraphics[width=0.46\textwidth]{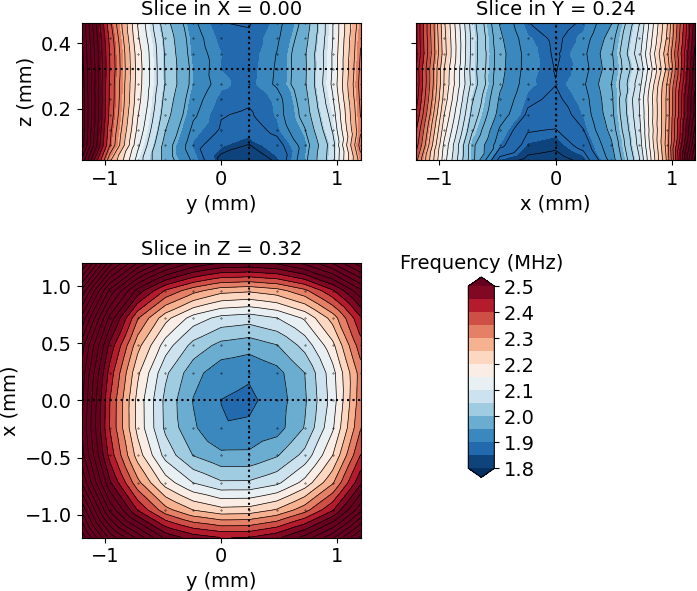}
    \caption{Measured Rabi frequency distribution maps for a circular polarized excited antenna at 2.8 W total input power. The excitation MW frequency was 2.81 GHz, corresponding to the $m_{s} = 0 \Longleftrightarrow -1$ transition at an external magnetic field of $B_{z} \cong 2$ mT.}
    \label{fig:Rabi_map-CircPol}
\end{figure}

The resulting Rabi frequency maps are illustrated in \autoref{fig:Rabi_map-LinearPol} and \ref{fig:Rabi_map-CircPol}. A uniform sampling grid was used with 0.05 mm x 0.05 mm resolution in the xy-direction and 0.02 mm in the z-direction. The maps exhibit good agreement with the magnetic fields obtained from the simulations of (see \autoref{fig:H_map_simul-LinearPol} and \ref{fig:H_map_simul-CircPol}). Among the observed frequencies, the Rabi frequency recorded in the symmetry center of the contour plots is 1.86 MHz for circular polarization and 1.55 MHz for linear polarization. These Rabi frequencies correspond to root mean square ($H_{RMS}$) fields of 22.3 and 22.6 A/m and per watt (total input power) for the circular and linear modes respectively, and are in good agreement with the 22.8 A/m predicted by the simulations. 

\subsection{Power test and temperature dependence}
\begin{figure}[hb]
     \centering
     \begin{tabular}{@{}c@{}c}
     \includegraphics[width=0.23\textwidth]
     {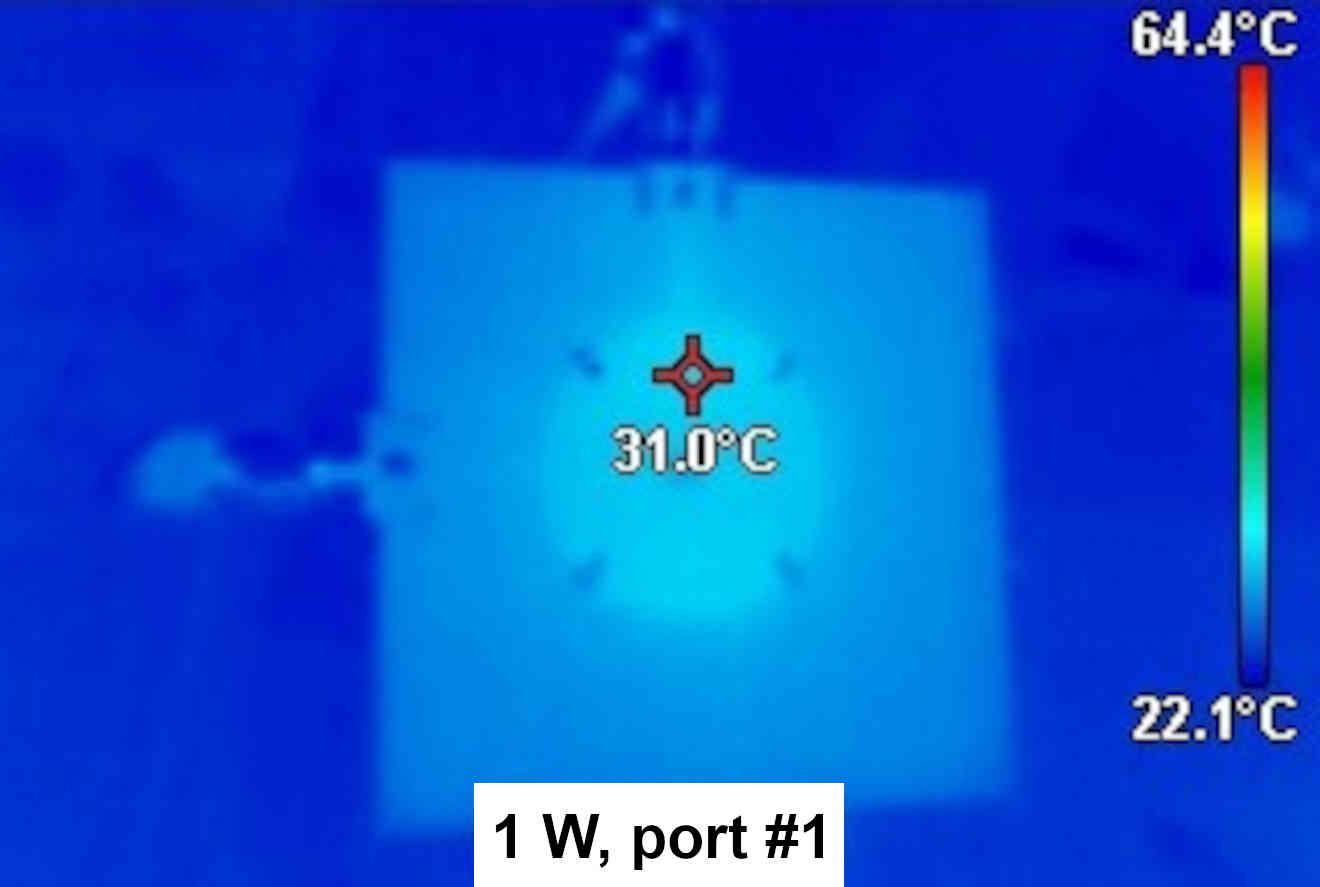}
     \end{tabular}
     \begin{tabular}{@{}c@{}}
     \includegraphics[width=0.23\textwidth]
     {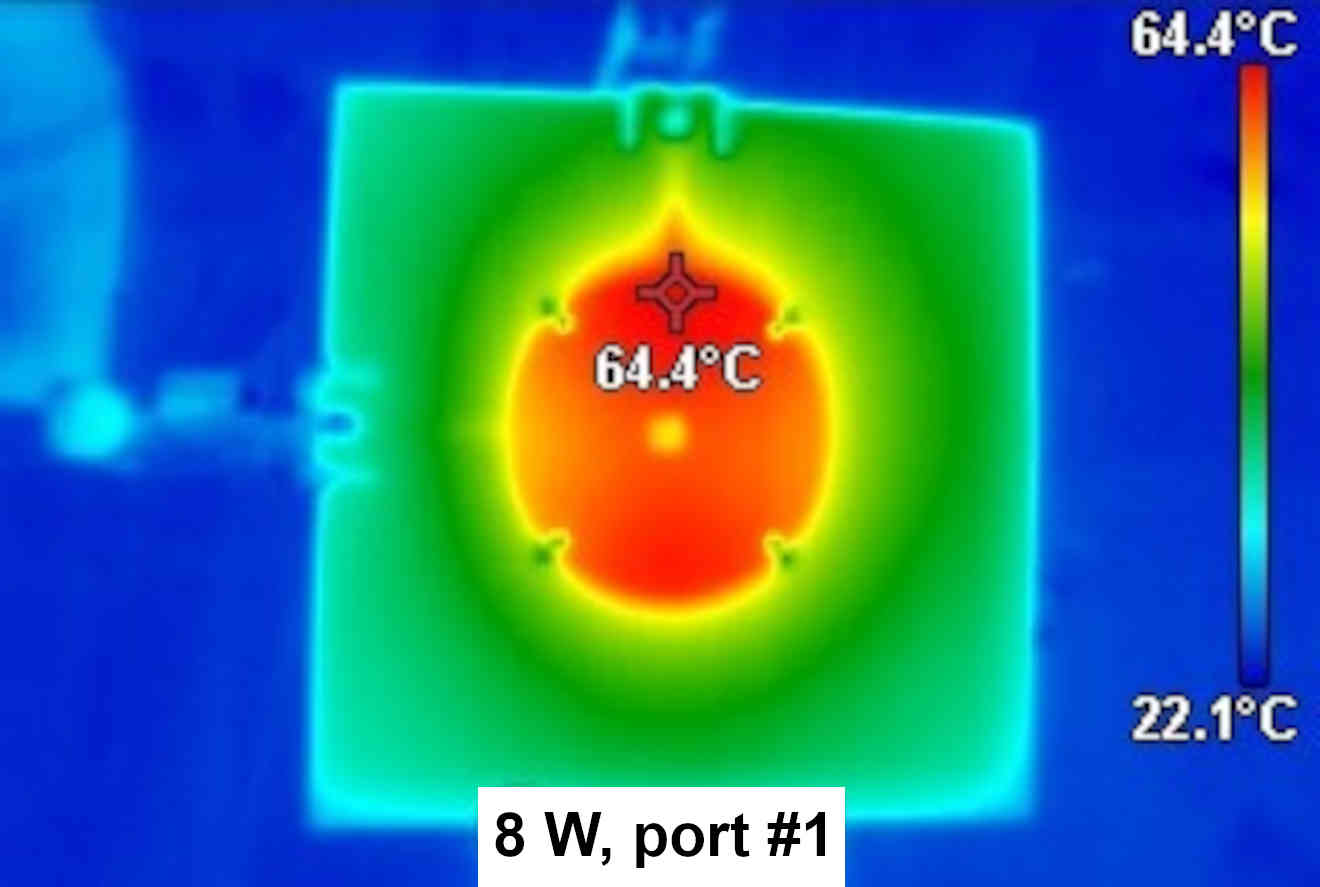}
     \end{tabular}
     \caption{The heat distribution in the antenna when put under 1 (left) and 8 (right) watts of continuous wave MW power on a single port.}
     \label{fig:temperature_graphs}
\end{figure}

Increasing the power delivered in the center by feeding more power could cause damage to the antenna due to heating. To test this, the heat distribution is measured with an infrared thermometer. \autoref{fig:temperature_graphs} shows that the antenna can handle the maximum output power of the amplifier used, which is limited to $8$W after transmission line losses. This temperature is below the safe human handling temperature limits set for consumer electronics devices \cite{InternationalElectrotechnicalCommissionIEC2023Audio/videoRequirements}. The antenna also shows a linear behaviour with the square root of the power as can be seen in \autoref{fig:rabi_freq_power}. 

Besides, it was observed that the temperature increase generated a frequency shift in the resonant behaviour of the antenna. \autoref{fig:s_parameters_ext_conditions} shows that the resonant frequency decreased by 16 MHz when increasing the temperature of the antenna from $22^{\circ} C$ to $40^{\circ} C$.

\section{Discussion}
Our experience shows that, while the new geometry is effective, the built antennas tend to have a lower resonant frequency than predicted by simulations. We distinguish three possible factors: the dielectric constant might differ, fabrication inaccuracies might be present or spurious electrical capacitances that were not considered might exist. The shift is about 55 MHz in the manufactured prototypes, but note that this may vary from producer to producer. Given this tendency to shift, and the limitation that capacitor based tuning can only lower the resonance frequency, the authors recommend scaling down the outer diameter of the copper patch aiming for a higher resonant frequency than the desired one.

The results presented herein question the need to add 2 additional dummy ports (a total of 4 ports) to the PCB to achieve high homogeneity volumes by geometric symmetry, as found earlier \cite{Herrmann2016polarization-Diamond}. Arguably, the addition of the two extra ports may reduce efficiency due to power leaking, which may partially justify that the design here proposed doubles the power efficiency as compared with the work of Herrmann et al. \cite{Herrmann2016polarization-Diamond}. Note that they made the diamond hole size smaller and covered the ground plane hole with a copper socket, modifications which should have significantly increased the field strength.

Selecting another material with a higher dielectric constant could enable creating higher magnetic fields and quality factors, as demonstrated in \cite{Yaroshenko2020CircularlyDiamond}. It should be noted that materials with higher dielectric constants come at higher cost and are less available. Furthermore, although these materials can in principle enhance the performance, the antenna also becomes more susceptible to manufacturing and material imperfections.

The central hole in the antenna offers flexibility for adjustment of the diamond along the axis of the hole. For example, the diamond can be located at the height of the ground plane for applications benefiting from microscope objectives with short ($>$0.6 mm) working distances. However, the highest field strength versus homogeneity is achieved when the diamond is centered along the hole axis. In this configuration, the field inhomogeneity is below 7\% within a volume of $1 \text{mm}^3$. This makes our configuration particularly suited to both harvest the averaging power of large pools of NV centers as well as those at moderate Rabi frequencies ($>$5 MHz), such as in dynamic nuclear hyperpolarization (DNP) and microfluidic NMR applications. 

Importantly, the high homogeneity volume in this antenna is cylindrical and covers a 0.5 mm height, which is a standard size of commercially available diamonds. Furthermore, this cylindrical geometry enables matching seamlessly the volume geometry addressed with the green pump beam to that of the low inhomogeneity MW field region. Our configuration largely improves the coherence in the response of the large pool of NVs, which is paramount for bulk sensing in applications such as magnetometry with bulk diamonds.

The $\sim$2 ns ringing time of the antenna does not impose a hazardous limitation as it is very short compared to those required for popular $\pi$ and $\pi$/2 pulses, which are more than an order of magnitude longer for the Rabi frequencies generated by this antenna. For example, achieving a $\pi/2$ pulse in 50 ns would require $<$20 W input power in the circularly polarized mode, which is enabled by the good linearity (\autoref{fig:rabi_freq_power}). The power can be safely increased to a maximum of $8$ W in continuous-wave mode without additional cooling (see \autoref{fig:temperature_graphs}). The structure could tolerate higher powers in pulsed mode to increase the MW field strength even further. Such higher powers could be implemented by using a pair of power amplifiers like the one used in this work to drive each port in this work. This feature would also facilitate fine-tuning each port's phase and amplitude to minimize the ellipticity in the induced MW field.

The direction of the circular polarization could be dynamically switched or adjusted, for example, by driving independently each port with a two channel arbitrary waveform generator or, for lower budget setups, by swapping the order of the transmission paths with fast switches. It is important to note the ellipticity of the field maps for linear polarization feeds. This ellipticity introduces a linear polarization component that limits the degree of isolation achievable when trying to address the -1 and +1 spin states independently.

\section{Conclusion}
Our results show that our antenna design offers a good compromise with respect to other antennas used in NV centers measurements. Our design doubles the power efficiency, offers an improved optical access and, most importantly, is robust against the variability of external hardware components, in contradiction to the previous patch antenna \cite{Herrmann2016polarization-Diamond}. By an inhomogeneity of 13\% in a $1 \text{ mm}^3$ volume, the proposed resonant MW antenna can address large ensembles of NVs in bulk diamond. Furthermore, this volume of low inhomogeneity can be easily and efficiently addressed optically. This limits the signal acquisition to only the NV centers that are coherently driven, which is paramount in boosting sensing precision. Our antenna design is particularly appealing for sensing experiments that benefit from lifting the -1 and +1 spin degeneracy without forcing an external magnetic field, such as at natural earth field conditions at zero field. The antenna's versatility is enhanced by its capability to become broadband over a wide range of frequencies through the use of capacitors and/or varactors. Its homogeneity also allows for easy positioning and repositioning without sacrificing accuracy on the flip angle. All these properties make this antenna a versatile piece of instrumentation for a wide range of high-end NV based quantum technologies.

\section*{Acknowledgment}
This project has been supported in part by European Union’s Marie Skłodowska-Curie project nr. 101030868, Erasmus+ exchange program from the European Union, the project I+D+i LINEAS ESTRATÉGICAS PLEC2021 from the Spanish Government (European NextGeneration funds), the Basque Government through the direct Subsidy for the Strengthening of the Quantum Technology Laboratory -PARTIDA DIRECTA 2023 action nr. 2023/01135 and the ELKARTEK 2022 program under Grant KK-2022/00062, the Programa Gipuzkoa Quantum 2023 -QUAN-000027-01 grant from the Diputación Foral De Gipuzkoa and the Plataforma de Tecnologías Cuánticas PTI-01 from CSIC. J.C. acknowledges the Ramón y Cajal (RYC2018-025197-I) research fellowship, the financial support from Spanish Government via the Nanoscale NMR and complex systems (PID2021-126694NB-C21) project, and the Basque Government grant IT1470-22.

\bibliographystyle{quantum}

\bibliography{referencesR}

\begin{thebibliography}{10}

\bibitem{Schirhagl2014Nitrogen-VacancyBiologyb}
Romana Schirhagl, Kevin Chang, Michael Loretz, and Christian~L. Degen.
\newblock ``{Nitrogen-Vacancy Centers in Diamond: Nanoscale Sensors for Physics
  and Biology}''.
\newblock
  \href{https://dx.doi.org/10.1146/annurev-physchem-040513-103659}{Annual
  Review of Physical Chemistry {\bf 65}, 83--105}~(2014).

\bibitem{Budakian2023RoadmapImaging}
Raffi Budakian, Amit Finkler, Alexander Eichler, Martino Poggio, Christian~L
  Degen, Sahand Tabatabaei, Inhee Lee, P~Chris Hammel, Eugene~S Polzik, Tim~H
  Taminiau, et~al.
\newblock ``Roadmap on nanoscale magnetic resonance imaging''~(2023).
\newblock
  url:~\href{http://arxiv.org/abs/2312.08841}{http://arxiv.org/abs/2312.08841}.

\bibitem{Degen2016QuantumSensing}
C.~L. Degen, F.~Reinhard, and P.~Cappellaro.
\newblock ``{Quantum sensing}''.
\newblock \href{https://dx.doi.org/10.1103/RevModPhys.89.035002}{Reviews of
  Modern Physics{\bf 89}}~(2016).

\bibitem{Doherty2013TheDiamond}
Marcus~W. Doherty, Neil~B. Manson, Paul Delaney, Fedor Jelezko, Jörg
  Wrachtrup, and Lloyd~C.L. Hollenberg.
\newblock ``{The nitrogen-vacancy colour centre in diamond}''.
\newblock \href{https://dx.doi.org/10.1016/J.PHYSREP.2013.02.001}{Physics
  Reports {\bf 528}, 1--45}~(2013).

\bibitem{Acosta2010OpticalDiamond}
V.~M. Acosta, A.~Jarmola, E.~Bauch, and D.~Budker.
\newblock ``{Optical properties of the nitrogen-vacancy singlet levels in
  diamond}''.
\newblock \href{https://dx.doi.org/10.1103/PhysRevB.82.201202}{Physical Review
  B - Condensed Matter and Materials Physics{\bf 82}}~(2010).

\bibitem{Gali2019AbDiamond}
Ádám Gali.
\newblock ``{Ab initio theory of the nitrogen-vacancy center in
  diamond}''~(2019).

\bibitem{Lenz2021MagneticCenter}
Till Lenz, Arne Wickenbrock, Fedor Jelezko, Gopalakrishnan Balasubramanian, and
  Dmitry Budker.
\newblock ``{Magnetic sensing at zero field with a single nitrogen-vacancy
  center}''.
\newblock \href{https://dx.doi.org/10.1088/2058-9565/abffbd}{Quantum Science
  and Technology {\bf 6}, 034006}~(2021).

\bibitem{Vetter2022Zero-Centers}
Philipp~J. Vetter, Alastair Marshall, Genko~T. Genov, Tim~F. Weiss, Nico
  Striegler, Eva~F. Gro{\ss}mann, Santiago Oviedo-Casado, Javier Cerrillo,
  Javier Prior, Philipp Neumann, and Fedor Jelezko.
\newblock ``{Zero- and Low-Field Sensing with Nitrogen-Vacancy Centers}''.
\newblock \href{https://dx.doi.org/10.1103/PhysRevApplied.17.044028}{Physical
  Review Applied{\bf 17}}~(2022).

\bibitem{Alegre2007polarization-selectiveDiamond}
Thiago~P.Mayer Alegre, Charles Santori, Gilberto Medeiros-Ribeiro, and
  Raymond~G. Beausoleil.
\newblock ``{Polarization-selective excitation of nitrogen vacancy centers in
  diamond}''.
\newblock \href{https://dx.doi.org/10.1103/PhysRevB.76.165205}{Physical Review
  B - Condensed Matter and Materials Physics{\bf 76}}~(2007).

\bibitem{Schloss2018SimultaneousSpins}
Jennifer~M. Schloss, John~F. Barry, Matthew~J. Turner, and Ronald~L. Walsworth.
\newblock ``{Simultaneous Broadband Vector Magnetometry Using Solid-State
  Spins}''.
\newblock \href{https://dx.doi.org/10.1103/PhysRevApplied.10.034044}{Physical
  Review Applied {\bf 10}, 034044}~(2018).

\bibitem{LeSage2013OpticalCells}
D.~Le~Sage, K.~Arai, D.~R. Glenn, S.~J. DeVience, L.~M. Pham, L.~Rahn-Lee,
  M.~D. Lukin, A.~Yacoby, A.~Komeili, and R.~L. Walsworth.
\newblock ``{Optical magnetic imaging of living cells}''.
\newblock \href{https://dx.doi.org/10.1038/nature12072}{Nature {\bf 496},
  486--489}~(2013).

\bibitem{Taylor2008High-sensitivityResolution}
J.~M. Taylor, P.~Cappellaro, L.~Childress, L.~Jiang, D.~Budker, P.~R. Hemmer,
  A.~Yacoby, R.~Walsworth, and M.~D. Lukin.
\newblock ``{High-sensitivity diamond magnetometer with nanoscale
  resolution}''.
\newblock \href{https://dx.doi.org/10.1038/nphys1075}{Nature Physics {\bf 4},
  810--816}~(2008).

\bibitem{Glenn2018High-resolutionSensor}
David~R. Glenn, Dominik~B. Bucher, Junghyun Lee, Mikhail~D. Lukin, Hongkun
  Park, and Ronald~L. Walsworth.
\newblock ``{High-resolution magnetic resonance spectroscopy using a
  Solid-State spin sensor}''.
\newblock \href{https://dx.doi.org/10.1038/nature25781}{Nature {\bf 555},
  351--354}~(2018).

\bibitem{Abe2018Tutorial:IN}
Eisuke Abe and Kento Sasaki.
\newblock ``{Tutorial: Magnetic resonance with nitrogen-vacancy centers in
  diamond-microwave engineering, materials science, and magnetometry}''.
\newblock \href{https://dx.doi.org/10.1063/1.5011231}{J. Appl. Phys {\bf 123},
  161101}~(2018).

\bibitem{Bucher2019QuantumSpectroscopy}
Dominik~B. Bucher, Diana~P.L. Aude~Craik, Mikael~P. Backlund, Matthew~J.
  Turner, Oren Ben~Dor, David~R. Glenn, and Ronald~L. Walsworth.
\newblock ``{Quantum diamond spectrometer for nanoscale NMR and ESR
  spectroscopy}''.
\newblock \href{https://dx.doi.org/10.1038/s41596-019-0201-3}{Nature Protocols
  {\bf 14}, 2707--2747}~(2019).

\bibitem{Wang2020IntegratedSensorsb}
Yuwan Wang, Yusong Liu, Hao Guo, Xingcheng Han, Anjiang Cai, Shengkun Li,
  Pengfei Zhao, and Jun Liu.
\newblock ``{Integrated microwave cavity and antenna to improve the sensitivity
  of diamond NV center spin-based sensors}''.
\newblock \href{https://dx.doi.org/10.35848/1882-0786/abbb3b}{Applied Physics
  Express {\bf 13}, 112002}~(2020).

\bibitem{Opaluch2021OptimizedApplications}
Oliver~Roman Opaluch, Nimba Oshnik, Richard Nelz, and Elke Neu.
\newblock ``{Optimized planar microwave antenna for nitrogen vacancy center
  based sensing applications}''.
\newblock \href{https://dx.doi.org/10.3390/nano11082108}{Nanomaterials{\bf
  11}}~(2021).

\bibitem{Sasaki2016BroadbandDiamond}
Kento Sasaki, Yasuaki Monnai, Soya Saijo, Ryushiro Fujita, Hideyuki Watanabe,
  Junko Ishi-Hayase, Kohei~M. Itoh, and Eisuke Abe.
\newblock ``{Broadband, large-area microwave antenna for optically detected
  magnetic resonance of nitrogen-vacancy centers in diamond}''.
\newblock \href{https://dx.doi.org/10.1063/1.4952418}{Review of Scientific
  Instruments {\bf 87}, 053904}~(2016).

\bibitem{Bayat2014EfficientResonators}
Khadijeh Bayat, Jennifer Choy, Mahdi Farrokh~Baroughi, Srujan Meesala, and
  Marko Loncar.
\newblock ``{Efficient, Uniform, and Large Area Microwave Magnetic Coupling to
  NV Centers in Diamond Using Double Split-Ring Resonators}''.
\newblock \href{https://dx.doi.org/10.1021/nl404072s}{Nano Lett {\bf 14},
  1208--1213}~(2014).

\bibitem{Chen2018Large-areaDiamond}
Yulei Chen, Hao Guo, Wangwang Li, Dajin Wu, Qiang Zhu, Binbin Zhao, Lei Wang,
  Yang Zhang, Rui Zhao, Wenyao Liu, Fangfang Du, Jun Tang, and Jun Liu.
\newblock ``{Large-area, tridimensional uniform microwave antenna for quantum
  sensing based on nitrogen-vacancy centers in diamond}''.
\newblock \href{https://dx.doi.org/10.7567/APEX.11.123001}{Applied Physics
  Express{\bf 11}}~(2018).

\bibitem{London2014StrongFields}
P.~London, P.~Balasubramanian, B.~Naydenov, L.~P. McGuinness, and F.~Jelezko.
\newblock ``{Strong driving of a single spin using arbitrarily polarized
  fields}''.
\newblock \href{https://dx.doi.org/10.1103/PhysRevA.90.012302}{Physical Review
  A - Atomic, Molecular, and Optical Physics{\bf 90}}~(2014).

\bibitem{Mrozek2015CircularlyDiamonds}
M.~Mr{\'{o}}zek, J.~Mlynarczyk, D.~S. Rudnicki, and W.~Gawlik.
\newblock ``{Circularly polarized microwaves for magnetic resonance study in
  the GHz range: Application to nitrogen-vacancy in diamonds}''.
\newblock \href{https://dx.doi.org/10.1063/1.4923252}{Applied Physics
  Letters{\bf 107}}~(2015).

\bibitem{MayerAlegre2007Microstrippolarization}
T.~P. Mayer~Alegre, A.~C. Torrezan, and G.~Medeiros-Ribeiro.
\newblock ``{Microstrip resonator for microwaves with controllable
  polarization}''.
\newblock \href{https://dx.doi.org/10.1063/1.2809372}{Applied Physics
  Letters{\bf 91}}~(2007).

\bibitem{Staacke2020MethodManipulation}
Robert Staacke, Roger John, Max Knei{\ss}, Christian Osterkamp, Séverine
  Diziain, Fedor Jelezko, Marius Grundmann, and Jan Meijer.
\newblock ``{Method of full polarization control of microwave fields in a
  scalable transparent structure for spin manipulation}''.
\newblock \href{https://dx.doi.org/10.1063/5.0030262}{Journal of Applied
  Physics{\bf 128}}~(2020).

\bibitem{Herrmann2016polarization-Diamond}
Johannes Herrmann, Marc~A. Appleton, Kento Sasaki, Yasuaki Monnai, Tokuyuki
  Teraji, Kohei~M. Itoh, and Eisuke Abe.
\newblock ``{Polarization- and frequency-tunable microwave circuit for
  selective excitation of nitrogen-vacancy spins in diamond}''.
\newblock \href{https://dx.doi.org/10.1063/1.4967378}{Applied Physics Letters
  {\bf 109}, 183111}~(2016).

\bibitem{Yaroshenko2020CircularlyDiamond}
Vitaly Yaroshenko, Vladimir Soshenko, Vadim Vorobyov, Stepan Bolshedvorskii,
  Elizaveta Nenasheva, Igor Kotel'nikov, Alexey Akimov, and Polina Kapitanova.
\newblock ``{Circularly polarized microwave antenna for nitrogen vacancy
  centers in diamond}''.
\newblock \href{https://dx.doi.org/10.1063/1.5129863}{Review of Scientific
  Instruments{\bf 91}}~(2020).

\bibitem{repoQuadMWPatchAntenna2024}
Open source PCB~repository.
\newblock \url{https://doi.org/10.5281/zenodo.10529652}.
\newblock Accessed: 2024-01-18.

\bibitem{Balanis2016AntennaDesign}
C~A Balanis.
\newblock ``{Antenna Theory: Analysis and Design}''.
\newblock Wiley. ~(2016).
\newblock
  url:~\href{https://books.google.nl/books?id=iFEBCgAAQBAJ}{books.google.nl/books?id=iFEBCgAAQBAJ}.

\bibitem{QfactorMeasurement2020}
Ming Zhang and Nicolas Llaser.
\newblock ``Design of analog and mixed circuits for resonator's q-factor
  measurement''.
\newblock In 2020 27th IEEE International Conference on Electronics, Circuits
  and Systems (ICECS).
\newblock \href{https://dx.doi.org/10.1109/ICECS49266.2020.9294784}{Pages
  1--4}.
\newblock ~(2020).

\bibitem{InternationalElectrotechnicalCommissionIEC2023Audio/videoRequirements}
{International Electrotechnical Commission (IEC)}.
\newblock ``{Audio/video, information and communication technology equipment -
  Part 1: Safety requirements}''~(2023).

\end{thebibliography}

\end{document}